\documentclass{article}

\usepackage[preprint]{neurips_2024}

\usepackage[utf8]{inputenc} % allow utf-8 input
\usepackage[T1]{fontenc}    % use 8-bit T1 fonts
\PassOptionsToPackage{hyphens}{url}  % allow hyphenation in URLs
\usepackage{hyperref}       % hyperlinks
\usepackage{url}            % simple URL typesetting
\usepackage{booktabs}       % professional-quality tables
\usepackage{amsfonts}       % blackboard math symbols
\usepackage{nicefrac}       % compact symbols for 1/2, etc.
\usepackage{microtype}      % microtypography
\usepackage{xcolor}         % colors
\usepackage{graphicx}       % includegraphics
\usepackage{listings}       % code listings
\usepackage{acro}           % acronyms
\usepackage{setspace}       % line spacing

\usepackage{xspace}
\newcommand{\datasetname}{PoliTok-DE\xspace}

\title{\datasetname: A Multimodal Dataset of Political TikToks and Deletions From Germany}

\author{%
  Tomas Ruiz$^{1}$ \qquad
  Andreas Nanz$^{2}$ \qquad
  Ursula Kristin Schmid$^{1}$ \\
  \textbf{Carsten Schwemmer}$^{1}$ \qquad
  \textbf{Yannis Theocharis}$^{2}$ \qquad
  \textbf{Diana Rieger}$^{1}$ \\
  $^{1}$ Ludwig Maximilian University of Munich\\
  $^{2}$ Technical University of Munich\\
  \texttt{\{t.ruiz,carsten.schwemmer\}@lmu.de}\\
  \texttt{\{ursula.schmid,diana.rieger\}@ifkw.lmu.de}\\
  \texttt{andreas.nanz@tum.de} \qquad
  \texttt{yannis.theocharis@hfp.tum.de}
}

\DeclareAcronym{afd}{
  short = AfD ,
  long = Alternative für Deutschland
}

\begin{document}

\onehalfspacing

\maketitle

\begin{abstract}
  We present \datasetname, a large-scale multimodal dataset (video, audio, images, text) of TikTok posts from two German elections: the 2024 Saxony state election and the 2025 German federal election.
  The corpus contains over 930{,}000 posts, of which over 330{,}000 were later deleted from the platform (18.7\% of Saxony posts, 39.7\% of federal posts).
  In the federal-election collection, about two thirds of the deletions were creator withdrawals, and the platform-deletion rate we computed was 13.0\% of all posts, more than an order of magnitude (14--19$\times$) above the platform-wide rate TikTok reported.
  Posts were identified via the TikTok research API and complemented with web scraping to retrieve full multimodal media and metadata.
  \datasetname supports social science research across substantive and methodological agendas: substantive work on intolerance and political communication, and methodological work on platform policies around deleted content and qualitative-quantitative multimodal research.
  To illustrate, we report a case study on intolerance and entertainment in an annotated subset of deleted posts: about one in five posts conveyed intolerance and a majority conveyed humor.
\end{abstract}

\begin{center}
  \href{https://huggingface.co/datasets/tomasruiz/PoliTok-DE}{%
    \raisebox{-0.3\height}{\includegraphics[height=1.4em]{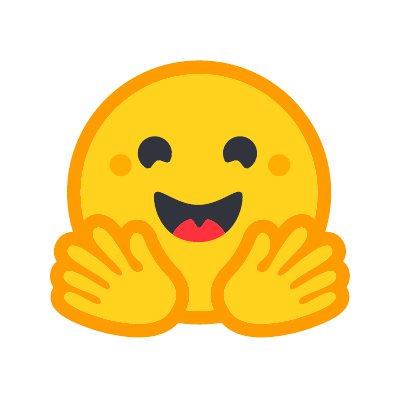}}\hspace{0.4em}\texttt{huggingface.co/datasets/tomasruiz/PoliTok-DE}}
\end{center}

\begin{figure}[h]
  \centering
  \includegraphics[width=0.8\linewidth]{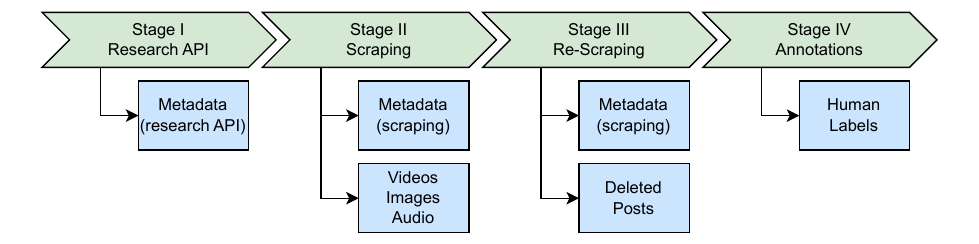}
  \caption{\textbf{Data collection process}. In green are the stages, and in blue the types of data collected across modalities (video, audio, images, and text). There are two sources of metadata: the research API and web scraping. Human labels were collected for a subset of posts.}
  \label{fig:data_collection}
\end{figure}

\section{Introduction}
We present \datasetname, a dataset of TikTok posts from two German elections: the 2024 Saxony state election and the 2025 German federal election.
The dataset contains over 930{,}000 posts in total.
For each post, we collected (1) post media (video, audio, images), (2) post metadata (description, hashtags, engagement metrics, etc.), and (3) whether the post was later deleted from the platform.
We publicly release the dataset of TikTok post IDs on Hugging Face and provide code to retrieve media and metadata directly from TikTok (hydration code).

A central feature of \datasetname is that it records which posts were later deleted: something a one-time scrape misses entirely.
We find that a large share of political content disappears in the months after an election, for two very different reasons: (1) creators quietly withdrawing their own posts, and (2) the platform deleting content under its rules.
Creator withdrawals dominate (2:1 ratio) yet never appear in platform transparency reporting, while platform deletions are far more common than TikTok reports.

Turning to the content, the right-wing party \ac{afd} is the most-mentioned party in both elections, and it is even more prevalent among the deleted posts.
The federal campaign was also far more personalized: candidates Alice Weidel and Friedrich Merz were each mentioned more often than most parties, unlike anyone in the state election.

Beyond these findings, \datasetname supports a broad range of research: (i) platform policies around deleted content, (ii) political communication, (iii) entertainment and intolerance, and (iv) qualitative, quantitative, and mixed-methods multimodal analysis.
To show what such work can look like in practice, we include a case study on the interplay of entertainment and intolerance (\autoref{sec:case-study}).

Our contributions are:
\begin{itemize}
  \item A two-election multimodal corpus of over 930{,}000 posts, with post media, metadata, and deletion snapshots recorded at two different time horizons, along with a description of our data collection pipeline (research API and web scraping) and hydration code.
  \item The deletion finding: about a third of deletions are platform takedowns, which run far above the rate TikTok reports, while about two thirds are creators withdrawing their own posts, a share invisible in platform transparency reports.
  \item A case study using an annotated subset of \datasetname to examine the co-occurrence of intolerance and entertainment.
\end{itemize}

The remainder of the paper is organized as follows:
\autoref{sec:related-work} situates \datasetname within prior work, \autoref{sec:data-collection} describes our data collection process, and \autoref{sec:dataset-statistics} presents the resulting dataset and its statistics.
\autoref{sec:case-study} then turns to the case study, and \autoref{sec:research-potential} concludes by discussing the broader research potential of the dataset.

\section{Related Work}
\label{sec:related-work}
Political TikTok is now an established research topic, and early work already showed that political messages on the platform combine text, sound, visuals, and platform-native interaction~\citep{medinaserrano2020dancing}.
Large-scale political TikTok datasets have mostly focused on the United States, including political-account collections, election and advertising datasets, elite-account studies, and work on toxicity and engagement~\citep{guinaudeau2022fifteen,papakyriakopoulos2022algorithms,pinto2024tracking,cheng2025politicalcontent,biswas2025toxicpolitics}.
These studies show TikTok's political relevance, but they are mainly U.S.-centered, account-centered, or based on metadata, transcripts, and other text-derived representations rather than full multimodal content.

German political TikTok research is also growing, but existing datasets cover narrow parts of the ecosystem: political influencers, German-speaking news outlets, politician and party accounts, or algorithmic exposure through sock-puppet audits~\citep{sehl2023politicalknowledge,mayer2025news,solovev2025tiktokrewards,wolfgram2024tiktok,tjaden2025feedleanright}.
These studies are close in setting, especially the German election audits, but they do not provide a broad keyword-based corpus that captures ordinary users, influencers, supporter networks, news outlets, and parties together in one corpus.

The data-access context further motivates such a resource. API access to social media data has become increasingly restricted~\citep{freelon2024postapi}, and TikTok's Research API has documented gaps and metric inconsistencies~\citep{pearson2025beyond,entrenaserrano2025tiktoksresearchapiproblems}.
These limits have led researchers to use alternatives such as data donation, screen tracking~\citep{ohme2024digitaltrace}, and ID-based sampling~\citep{steel2025justhourtiktokreverseengineering}.
We nevertheless use API-based collection because it supports broad keyword-based retrieval over long multi-month election windows.
Closest in genre, \citet{barrie2025enriched} document a multimodal election dataset, though for X rather than a video-native platform.

\datasetname adds to the body of literature by providing a large German political TikTok dataset that combines a broad election-focused keyword collection over several months with multimodal content, human annotations, and deletion tracking.
We follow best practices for documenting the collection process, dataset composition, limitations, and access conditions~\citep{gebru2021datasheets}.

\section{Data Collection}
\label{sec:data-collection}
Our dataset comprises two collections of multimodal TikTok posts, one for each of two German elections:
the Saxony state election of September 2024 and the federal election (Bundestagswahl) of February 2025.
The dataset was gathered using a four-stage pipeline, shown in \autoref{fig:data_collection}:
we found election-related posts using the TikTok research API (stage I), scraped their full media from the TikTok website (stage II), later determined which posts had been deleted from the platform (stage III), and annotated a sample of the posts (stage IV).
We describe each stage below, and then introduce each collection.

\paragraph{Stage I: Research API}
The TikTok research API provides a way to search for posts using \textit{keywords}.
For each collection, we formulated a list of keywords related to the election (\emph{e.g.}\ \textit{afd}, \textit{bundestag}) and queried the research API (the full keyword lists are in \autoref{sec:research-api-keywords}).
Besides the keywords, we also restricted the query to content posted in Germany (\textit{region=DE}).
During the collection window, we collected posts daily.
According to TikTok documentation\footnote{\url{https://developers.tiktok.com/doc/research-api-faq}},
the research API needs 48 hours (2 days) to mirror the content from the website.
The content that is deleted from TikTok during this time window never makes it to the research API.
Our own experiments showed a larger time lag,
so we collected the data for each day 96 hours (4 days) later.
The research API provided post IDs and metadata, \emph{i.e.}\ post description, engagement metrics, and more.
However, the research API does \textit{not} provide the post media, such as video, images, and audio, so we retrieved them in the next stage.
We also identified two undocumented behaviors of the research API, related to how keywords are matched and the random-sampling mode, which we document in \autoref{sec:api-behaviors}.

\paragraph{Stage II: Scraping}
In the second stage, we scraped the post multimodal media (video, audio, images) directly from the TikTok website.
This happened daily, directly after stage I.
Besides the post media, the second stage also provided post metadata, \emph{e.g.}\ engagement metrics, that could be compared to the metadata provided by the research API. 

\paragraph{Stage III: Find Deleted Posts}
In the third stage, we determined which posts were deleted from the platform by re-scraping the posts from the website.
We performed this re-scraping once or repeatedly after the end of the collection window.
The deletions can have different reasons, \emph{e.g.}\ platform policy violations, as well as users deleting their own posts or switching them to private mode, where they are no longer visible to the public.

\paragraph{Stage IV: Annotations}
In stage four, we sampled posts from the dataset and annotated them for our case study (see \autoref{sec:case-study}).

\subsection{Collections}
\paragraph{Collection 1: The 2024 Saxony State Election}
\label{sec:saxony-collection}
The first collection covers the Saxony state election (Landtagswahl) of 01.09.2024.
The keyword list covers the political parties, leading politicians, and mentions of the state of Saxony (full list in \autoref{sec:research-api-keywords}).
We collected posts published between 01.07.2024 and 30.11.2024, \emph{i.e.}\ starting 2 months before and ending 3 months after the election, for a total of 195{,}373 posts.
We re-scraped the posts (stage III) repeatedly, between 1 and 4.5 months after the election (\autoref{sec:content-removal}).
A sample of this collection was annotated (stage IV) for our case study (see \autoref{sec:case-study}).

\paragraph{Collection 2: The 2025 Federal Election}
\label{sec:btw-collection}
The second collection covers the German federal election (Bundestagswahl) of 23.02.2025, a snap election following the collapse of the governing coalition in November 2024~\citep{euronews_coalition_2024}.
The keyword list covers the political parties, leading politicians including the chancellor candidates, and two issues that were salient in the campaign: migration and the Russia--Ukraine war (full list in \autoref{sec:research-api-keywords}).
The release covers posts published between 13.01.2025 and 30.04.2025 (\textasciitilde{}6 weeks before and \textasciitilde{}9.5 weeks after the election).
Data collection before 13.01.2025 was patchy, so we exclude it from the release.
In total, we collected 1{,}066{,}769 posts.
After removing posts that only matched false-positive keywords (see \autoref{sec:research-api-keywords}) and restricting to the collection window, the released collection contains \textbf{743{,}588 posts}, about 3.8 times the Saxony collection.
We re-scraped all released posts (stage III) about 16 months after the election (June 2026), compared to 1--4.5 months for Saxony, so the two collections capture deletion at a short and a long time horizon (\autoref{sec:content-removal}).
The annotation stage (stage IV) for this collection is currently underway.

\section{Dataset Statistics}
\label{sec:dataset-statistics}
\begin{figure}[h]
  \centering
  \includegraphics[width=0.8 \linewidth]{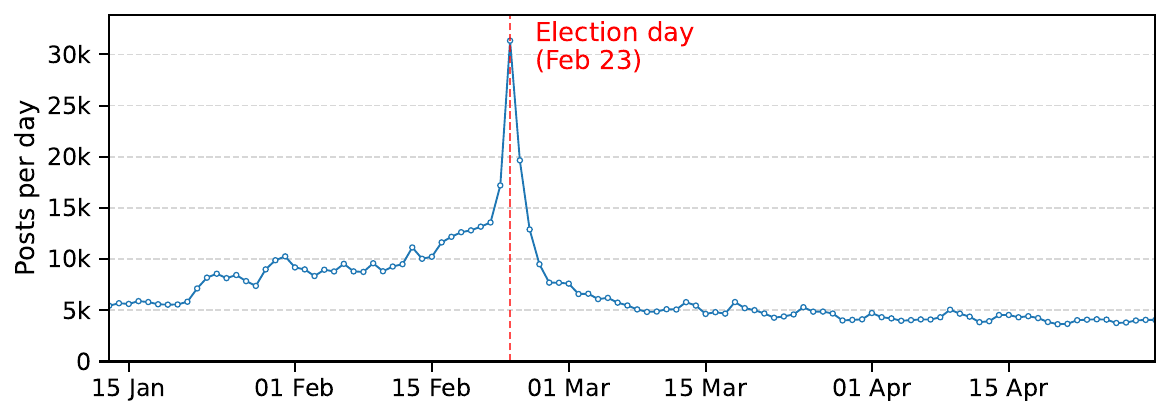}
  \caption{\textbf{Posts per Day (Federal Election 2025).} The daily volume peaks on election day (23.02.2025) and declines afterwards.}
  \label{fig:btw-posts-per-day}
\end{figure}

The dataset contains 938{,}961 posts in total: 195{,}373 posts in the Saxony collection and 743{,}588 posts in the federal-election collection (\autoref{tab:collections-overview}).
The federal-election collection is both larger and far more concentrated around its election day: its daily volume rises from about 5{,}000 posts in mid-January to a peak of 31{,}341 on election day (23.02.2025) and falls back to about 4{,}000 by April (\autoref{fig:btw-posts-per-day}).
We analyze the keywords prevalence in \autoref{sec:keyword-frequency} and the deletion rates in \autoref{sec:content-removal}.

\begin{table}[h]
  \caption{\textbf{Overview of the two collections.} The federal-election collection is about 3.8 times larger and far more concentrated around its election day.}
  \label{tab:collections-overview}
  \centering
  \begin{tabular}{lrr}
    \toprule
     & Saxony Election 2024 & Federal Election 2025 \\
    \midrule
    Election day & 01.09.2024 & 23.02.2025 \\
    Collection window & 01.07.2024--30.11.2024 & 13.01.2025--30.04.2025 \\
    Number of days & 153 & 108 \\
    Posts & 195{,}373 & 743{,}588 \\
    Average posts per day & 1{,}277 & 6{,}885 \\
    Peak posts per day & 3{,}147 & 31{,}341 \\
    Peak day & 08.11.2024 & 23.02.2025 (election day) \\
    \bottomrule
  \end{tabular}
\end{table}

\subsection{Keyword Frequency}
\label{sec:keyword-frequency}
\begin{figure}[h]
  \centering
  \includegraphics[width=\linewidth]{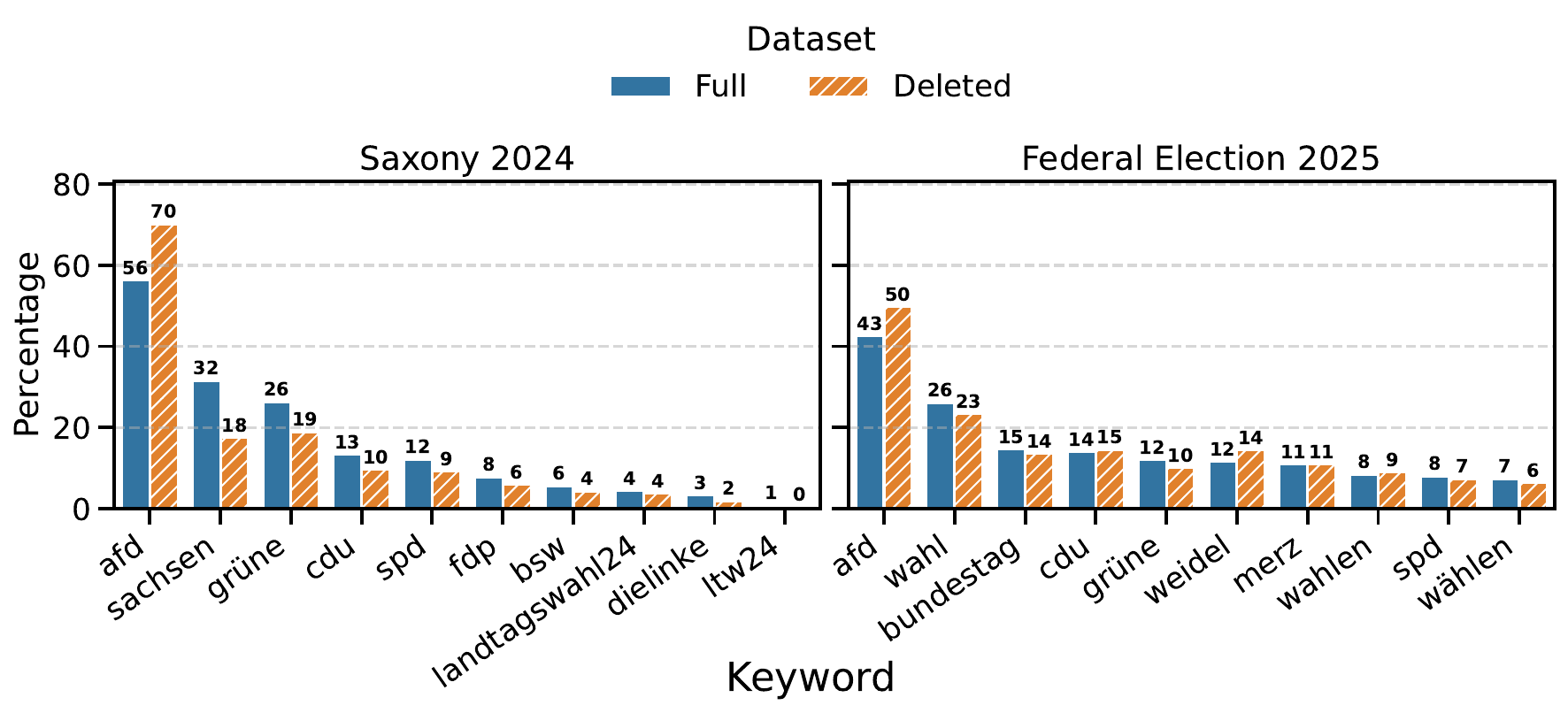}
  \caption{\textbf{Most Frequent Keywords.}
  \textbf{(left)} Saxony 2024: political parties are well represented, with the right-wing party \ac{afd} being mentioned in 56.3\% of all posts, and in 70.2\% of the deleted posts. The keyword ``sachsen'' (Saxony) was also prominent.
  \textbf{(right)} Federal Election 2025: the right-wing party \ac{afd} is again the most frequent party, mentioned in 42.6\% of all posts and 49.8\% of the deleted posts. The chancellor candidates Weidel and Merz are mentioned more often than most parties. The keyword \textit{wahl25} is shown as \textit{wahl}, the term it effectively matched (see \autoref{sec:api-behaviors}).
  Percentages shown on the bars are rounded to the nearest integer.}
  \label{fig:keyword-frequency}
\end{figure}

The top 10 keywords of each collection are shown in \autoref{fig:keyword-frequency}.
In both collections, the most frequent party by a large margin is the right-wing party \ac{afd}: it is mentioned in 56.3\% of the Saxony posts and in 42.6\% of the federal-election posts.
The second most frequent party differs: The Greens (26.4\%) in Saxony, and the CDU/CSU (14.4\%) followed by The Greens (12.0\%) in the federal election.
Within the deleted content, the proportion of posts mentioning the \ac{afd} is even higher: 70.2\% in Saxony (+13.9~pp) and 49.8\% in the federal election (+7.2~pp).
This shows a very strong presence of the \ac{afd} across our collected content, and most importantly in the deleted content.

The two collections differ markedly in how often candidates are mentioned, as shown in \autoref{tab:candidates}.
In the Saxony collection, even the most mentioned candidate, Michael Kretschmer, appears in only 0.27\% of the posts, and some candidate keywords had 0 mentions.
In the federal-election collection, the chancellor candidates Alice Weidel (11.6\%) and Friedrich Merz (10.9\%) were each mentioned in more posts than most parties.
This suggests that the discourse around the federal election was much more centered on the leading candidates than the discourse around the state election.

\begin{table}[h]
  \caption{\textbf{Top 5 candidates by share of posts mentioning them.} The discourse around the federal election was far more centered on candidates than the discourse around the state election.}
  \label{tab:candidates}
  \centering
  \begin{tabular}{lrlr}
    \toprule
    \multicolumn{2}{c}{Saxony Election 2024} & \multicolumn{2}{c}{Federal Election 2025} \\
    \cmidrule(r){1-2} \cmidrule(l){3-4}
    Candidate & \% & Candidate & \% \\
    \midrule
    Kretschmer (CDU) & 0.27 & Weidel (AfD) & 11.60 \\
    Köpping (SPD) & 0.06 & Merz (CDU) & 10.94 \\
    Urban (AfD) & 0.05 & Habeck (Grüne) & 2.96 \\
    Meier (Grüne) & $<$0.01 & Scholz (SPD) & 2.85 \\
    Hartmann (Linke) & $<$0.01 & Baerbock (Grüne) & 1.05 \\
    \bottomrule
  \end{tabular}
\end{table}

\subsection{Deletions}
\label{sec:content-removal}
We use \emph{deletion} to mean a post that was no longer retrievable when we re-scraped it, whether taken down by the creator (deleted or made private) or by the platform (moderation, geo-restriction).
The deleted share grows the longer we wait: in Saxony, where we re-scraped posts repeatedly between 1 and 4.5 months after the election, 36{,}617 posts (18.7\%) had been deleted by the last re-scrape.
In the federal election, which we re-scraped once 16 months after the election, 295{,}101 posts (39.7\%) had been deleted.

We group deletions into two kinds: (1) creator withdrawals, where the post was deleted or made private while the account stayed intact, and (2) platform deletions, where the post was under review, failed a moderation audit, was age-restricted, or was geo-blocked.
We tell the two apart from the structure of the status codes that TikTok returns for a deleted post: creator withdrawals appear as a single, isolated status code, whereas enforcement actions appear as a tightly co-occurring bundle of status codes (full details in \autoref{sec:removal-regimes}).
We call the share of posts in the second group our \textit{platform-deletion rate}.
For the federal-election collection, creator withdrawals account for 67.4\% of deleted posts, or 26.7\% of all posts (see \autoref{tab:btw-removal-regimes}).
\autoref{sec:removal-regimes} shows that the creator-withdrawal share grows with time after the election.
The platform-deletion rate is 13.0\% of all posts, which is far higher than the platform's own transparency reporting would suggest.

According to~\citet{tiktok_cge_2025q3}, fewer than 1\% of published videos are removed for Community Guidelines violations in any quarter (between 0.7\% and 0.9\% across the first three quarters of 2025), and about 94\% of those removals happen within the first 24 hours after a post is published.
Our platform-deletion rate is therefore more than an order of magnitude higher (13.0\% vs 0.7-0.9\%, or 14-19$\times$) than the platform-wide quarterly rate.
Creator withdrawals do not appear in such transparency reports at all, because they count only the videos the platform itself removes and not the posts that creators withdraw on their own.
Our finding establishes a very large gap between the platform's reported deletion rate and the actual deletion rate in political content.

\begin{table}[h]
  \caption{\textbf{Deletion Type (Federal Election 2025).} Of the 295{,}101 deleted federal-election posts, about two thirds are classified as creator withdrawals and about one third as platform deletions. Percentages are of deleted posts and of all 743{,}588 posts in the federal-election collection.}
  \label{tab:btw-removal-regimes}
  \centering
  \begin{tabular}{lrrr}
    \toprule
    Deletion Type & Posts & \% of deleted & \% of all \\
    \midrule
    Creator withdrawal & 198{,}790 & 67.4 & 26.7 \\
    Platform enforcement & 96{,}311 & 32.6 & 13.0 \\
    \midrule
    Total deleted & 295{,}101 & 100.0 & 39.7 \\
    \bottomrule
  \end{tabular}
\end{table}

\section{Research Potential}
\label{sec:research-potential}
The \datasetname dataset supports computational social science across substantive and methodological areas.

\paragraph{Substantive}
It allows studying entertainment in short-form political video (e.g., humor, music, visual overlays, pacing) using video, audio, images, and text. 
It also allows examining intolerance by identifying exclusionary rhetoric in descriptions, on-screen text, and audio, and comparing signals and cues across modalities. 
It enables studying political communication via party and issue mentions in descriptions/hashtags, audiovisual messaging choices, and associations with post popularity.

\paragraph{Methodological}
Because of the multimodality of the dataset, it supports rich and deep qualitative analysis, as well as large-scale quantitative coding/modeling. 
In particular, it enables mixed-methods approaches that leverage methods from both.
The presence of video and audio allows researchers to go beyond the text-only and still-image analysis prevalent in the literature.
The deletion snapshot enables comparisons between available and deleted posts by content characteristics, user attributes, and engagement.

\section{Case Study: Finding Toxic Entertainment}
\label{sec:case-study}
\paragraph{Goal} Our aim was to capture the political discourse on TikTok related to the 2024 state elections (Landtagswahlen)
in Saxony.
The right-wing party \ac{afd} had a strong presence in Saxony, finishing first in the 2024 state election
with over one-third of the vote~\citep{sachsen_landtagswahlen_2024}.
We were particularly interested in the combination of intolerance and entertainment: how humor and playful framing can co-occur with exclusionary messages.
Therefore, we focused our analysis on deleted posts to (a) hopefully increase the base rate of intolerant content and (b) include potential extreme cases (\emph{e.g.}\ potentially illegal content) that help us understand the phenomenon.

\paragraph{\ac{afd} on Social Media} The success of the right-wing party \ac{afd} on social media platforms is a matter of debate.
On the one hand, political observers (\citet{metzger_afd_2024}) and previous research (\cite{franke2022rechtsextremismus,hohner2024analyzing}) suggested that TikTok could become a core campaigning platform during the election, with a particularly strong presence of \ac{afd} content.
On the other hand, \citet{fielitz2024social} suggested that the \ac{afd}'s digital success is overstated and that the party depends on mainstream media and sympathetic influencers for reach.

\subsection{Annotations}
\label{sec:annotations}
\begin{figure}[h]
  \centering
  \includegraphics[width=\linewidth]{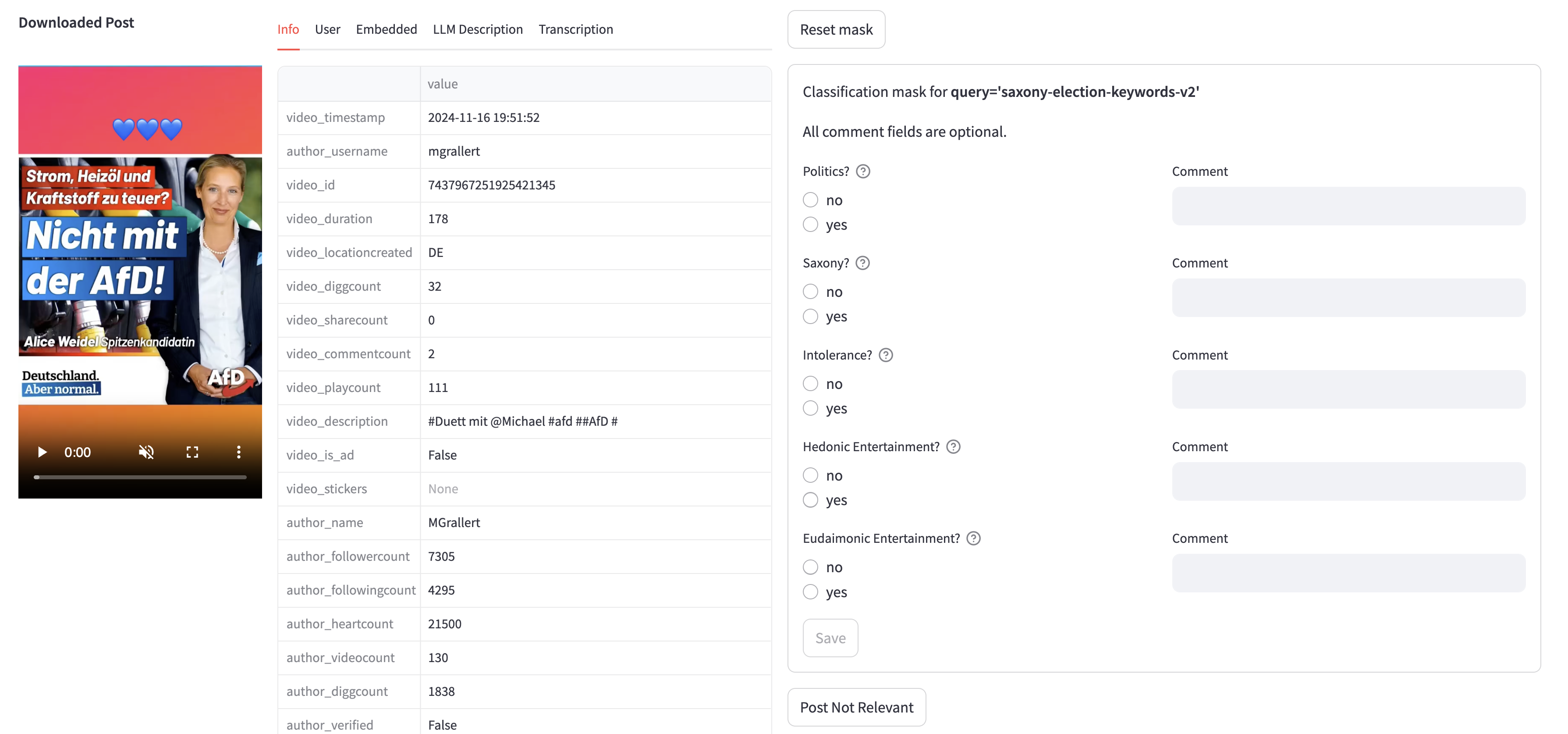}
  \caption{\textbf{Annotation UI}. The left half shows the multimodal post media (video, audio) and textual metadata. The right half shows the annotation mask. By hovering over the question mark, the annotator can see the full codebook for each question.}
  \label{fig:annotation-ui}
\end{figure}

We were interested in finding intolerant content on TikTok, but intolerant content makes up only a small fraction of all posts on the platform, even when our keywords were focused on political topics.
To increase the likelihood of finding intolerant content, we focused our annotation effort on deleted posts.
We annotated (1) whether the posts referred to political topics, 
(2) if they referred to Saxony, 
(3) if they conveyed intolerance, and (4) if they conveyed entertainment.
All annotations were binary (yes/no). Annotators were allowed to write an optional explanation in the comment field.
Each post was annotated by 1 to 4 annotators (mean = 2.57). The annotation mask is shown in \autoref{fig:annotation-ui}.
Annotators viewed the full multimodal post (video, audio, images) alongside textual metadata (like description and hashtags), and based their judgments on cues within and across modalities.
The full codebooks for all categories are provided in \autoref{sec:all-codebooks}.

\begin{itemize}
    \item \textbf{Politics:} Content that discusses or references policy, political events, actors, or institutions.
    \item \textbf{Saxony:} Content that refers to the federal state of Saxony, its politics, places, or public figures.
    \item \textbf{Intolerance:} Content that communicates the denial of an equal status with others and the refusal to accept different perspectives, cultures, or ideas \citep{rossini2022beyond}.
    \item \textbf{Entertainment:} We annotated two different types of entertainment, following \citet{oliver2011entertainment}.
    \begin{itemize}
        \item \textit{Hedonic entertainment:} Content that conveys fun, humor, excitement, and pleasure.
        \item \textit{Eudaimonic entertainment:} Content that conveys inspiring, touching, thought-provoking, and meaningful elements.
    \end{itemize}
\end{itemize}

\subsection{Results}

\paragraph{Annotations} \autoref{tab:inter-annotator-agreement} shows the response distribution (No/Yes) for each question.
We found that the majority of posts refer to politics, while only a minority refer to Saxony, or contain intolerance.
In terms of entertainment, most content conveyed hedonic entertainment, while a minority conveyed eudaimonic entertainment.
About 13\% of the posts were annotated as conveying both hedonic and eudaimonic entertainment.

\begin{table}[h]
  \caption{Response distribution (No/Yes) for each question, and inter-annotator agreement (Krippendorff's $\alpha$, higher values are better). The majority class is in \textbf{bold}.}
  \label{tab:inter-annotator-agreement}
  \centering
  \begin{tabular}{lcccc}
  \hline
  \textbf{Question} & \textbf{No (\%)} & \textbf{Yes (\%)} & \textbf{Agreement (\%)} & \textbf{Krip. Alpha} \\
  \hline
  Political & 14.3 & \textbf{85.7} & 94.4 & 0.81 \\
  Saxony & \textbf{80.8} & 19.2 & 90.8 & 0.74 \\
  Intolerance & \textbf{79.5} & 20.5 & 79.7 & 0.48 \\
  Hedonic Entertainment & 37.1 & \textbf{62.9} & 75.8 & 0.55 \\
  Eudaimonic Entertainment & \textbf{62.7} & 37.3 & 67.2 & 0.38 \\
  \hline
  \end{tabular}
\end{table}

\paragraph{Reliability} We calculated agreement percentage and Krippendorff's $\alpha$ for each question, as shown in \autoref{tab:inter-annotator-agreement}.
The $\alpha$ metric ranges from -1.0 to 1.0, with 1.0 being perfect agreement, and it controls for agreement by chance.
~\citet[p.241]{krippendorff2018content} recommended an $\alpha$ value of 0.80 in general, or at least 0.66 for tentative conclusions. 
Although our annotation agreement percentages are significantly above chance, they do not meet the recommended $\alpha$ values for the questions about intolerance and entertainment, even after iterating on the codebook and holding repeated annotator discussions.

The difficulty during annotation is related to the multimodality of content, and the \textit{implicit} forms of intolerance it enables.
Implicit intolerance makes annotations more difficult, and annotators are more likely to disagree about it~\citep{elsherief-etal-2021-latent}.
Multimodal content can encode implicit forms of intolerance across more modalities (visual, audio, and text), and in the \textit{interaction} of modalities.
Prior work analyzing multimodal intolerance has focused on meme-like images only~\citep{crisishatemm2023,fersini-etal-2022-semeval,pramanick-etal-2021-detecting,suryawanshi-etal-2020-multimodal}, which lack the multiple frames and audio cues of videos.

\section{Limitations}
For keyword prevalence, we used case-insensitive string matching in descriptions (including hashtags).
If a keyword is mentioned only in the audio or visuals of a post, it is not captured by this procedure, so prevalence metrics might be underestimated.
A deletion can have different causes (platform policy, user choice, privacy settings), so it does not necessarily imply policy-violating or harmful content.
The deletion status should be interpreted with this nuance in mind.
Since our two collections were re-scraped at different times, we do not compare their deletion rates directly, and instead show that deletion grows over time using the same Saxony posts re-scraped at 1, 3, and 4.5 months (\autoref{sec:removal-regimes}).
Our re-scrapes are snapshots, so posts with status code in-review could be later reinstated or be fully deleted.

\section{Ethical Considerations and Access}
With our data release, we seek to balance two goals:
advancing computational social science research and protecting the rights and privacy of TikTok users.
To avoid legal, copyright, and terms-of-service violations, we do not publish the raw content, such as videos, directly.
Instead, we release post identifiers and provide code that allows researchers to recollect the data from the platform (hydration code).
This approach supports reproducibility and broad access while reducing the risks of improper distribution.

We acknowledge that trade-offs remain.
Obtaining explicit consent from tens of thousands of TikTok users is infeasible, but our strategy still preserves a high degree of user privacy by avoiding direct redistribution of their content.
The post IDs are available through the Hugging Face website without review, but no media content is publicly released for any post.
Academic researchers can request access to the full stored media and metadata, and the authors will evaluate each request before granting access.
These safeguards do not fully resolve all concerns, but they significantly reduce risks while maintaining the integrity of the research process.
Despite the limitations, we believe our strategy strikes a positive overall balance, responsibly enabling impactful research, respecting platform policies, and minimizing potential harm to TikTok users.

\section{Conclusion}

We introduced \datasetname, a multimodal dataset of TikTok posts from two German elections, the 2024 Saxony state election and the 2025 federal election, that includes post media, platform metadata, and a record of deleted posts.
The dataset covers over 930{,}000 posts, of which over 330{,}000 (18.7\% of the Saxony posts and 39.7\% of the federal-election posts) were later deleted.
About two thirds of these deletions were creators withdrawing their own posts.
Platform enforcement accounted for the remaining third, but even on its own it ran far above the rate TikTok reports across its platform.
We described the data collection process including the research API, web scraping, and reported dataset statistics, including keyword frequencies and deletion prevalence.
We also presented a case study using an annotated subset to examine the co-occurrence of intolerance and entertainment.
Beyond this case study, \datasetname can support work in computational social science on entertainment and intolerance, political communication, deletion-related platform dynamics, and qualitative, quantitative, and mixed-methods multimodal research.
We encourage researchers to use the dataset for their research.
We publicly share the dataset of post IDs on Hugging Face (link in front-matter) together with hydration code,
and make deleted content available upon request for academic research purposes.

\begin{ack}
This research is funded by the Bavarian Research Institute for Digital Transformation (bidt),
and was conducted within the ToxicAInment research project.\footnote{\url{https://en.bidt.digital/research-project/using-ai-to-increase-resilience-against-toxicity-in-online-entertainment/}}
\end{ack}

\bibliographystyle{plainnat}
\bibliography{custom}

\appendix

\section{Research API Keywords}
\label{sec:research-api-keywords}
To query the research API, we use the \textbf{Query Videos} endpoint.\footnote{\url{https://open.tiktokapis.com/v2/research/video/query/}}
Our post-hoc keyword audit checks whether a keyword appears in the post description, including hashtags.
Matching is case-insensitive and uses substring matching, so \emph{e.g.}\ \textit{migrat} matches both \textit{Migration} and \textit{Migrationspolitik}.

\paragraph{Saxony 2024}
The set of keywords is below. It includes politicians, political parties, and mentions of the state of Saxony.

\textit{['Sachsen', 'ltw24', 'landtagswahl24', 'CDU', 'Afd', 'dielinke', 'grune', 'gruene', 'grüne', 'SPD', 'FDP', 'BSW', 'wahlkampfsachsen', 'Kretschmer', 'JörgUrban', 'joergurban', 'urbanjörg', 'urbanjoerg', 'SusanneSchaper', 'SchaperSusanne', 'StefanHartmann', 'HartmannStefan', 'KatjaMeier', 'MeierKatja', 'WolframGünter', 'GünterWolfram', 'WolframGuenter', 'GuenterWolfram', 'FranziskaSchubert', 'SchubertFranziska', 'PetraKöpping', 'KöppingPetra', 'petrakoepping', 'KoeppingPetra', 'RobertMalorny', 'MalornyRobert', 'SabineZimmermann', 'ZimmermannSabine']}

\paragraph{Federal Election 2025}
The set of keywords is below. It includes general election terms, political parties, leading politicians (including all chancellor candidates), and terms related to migration, climate and energy, and the Russia--Ukraine war.

\textit{['btw25', 'btw2025', 'bundestag', 'wahl25', 'wahlkampf', 'wählen', 'wahlen', 'kanzler', 'koalition', 'ampel', 'groko', 'cdu', 'csu', 'afd', 'dielinke', 'grune', 'gruene', 'grüne', 'bundnis90', 'spd', 'fdp', 'bsw', 'scholz', 'merz', 'habeck', 'lindner', 'weidel', 'reichinnek', 'vanaken', 'wagenknecht', 'jensspahn', 'söder', 'soder', 'baerbock', 'dobrindt', 'vonstorch', 'chrupalla', 'pistorius', 'wissing', 'kukies', 'faeser', 'özdemir', 'ozdemir', 'lauterbach', 'schulze', 'geywitz', 'höcke', 'hocke', 'schwerdtner', 'asyl', 'flüchtling', 'fluchtling', 'immigr', 'migrat', 'migrant', 'emigr', 'ausländer', 'auslander', 'ausländisch', 'auslandisch', 'zuwander', 'zugewander', 'einwander', 'eingewander', 'ausgewander', 'gastarbeiter', 'arbeitnehmerfreizügigkeit', 'arbeitnehmerfreizugigkeit', 'personenfreizügigkeit', 'personenfreizugigkeit', 'klima', 'energie', 'umwelt', 'wandel', 'wende', 'kohle', 'nachhaltigkeit', 'russland', 'russisch', 'ukrain', 'putin', 'sowjet', 'zelensk', 'selensk', 'kreml', 'minsk', 'nato', 'moskau', 'kiew', 'kiev', 'donezk', 'luhansk', 'donbas', 'bachmut', 'mariupol', 'kharkiv', 'krim', 'cherson', 'fsb', 'kgb', 'militärhilfe', 'militarhilfe', 'waffenlieferung', 'rüstungs', 'rustungs']}

\paragraph{Blacklisted Keywords (Federal Election 2025)}
After inspecting the collected content, we found that 10 of our keywords produced overwhelmingly false-positive matches, \emph{i.e.}\ they matched posts unrelated to the election.
The clearest example is \textit{btw25}: we intended it to match the common abbreviation for \textit{Bundestagswahl 2025}, but because the research API ignores trailing digits in keywords (see \autoref{sec:api-behaviors}), it actually matched the English abbreviation \textit{btw} (``by the way''), flooding the collection with unrelated English-language posts.
The remaining keywords are common German words that mostly matched non-political content (\emph{e.g.}\ \textit{wende} as part of \textit{anwenden}) or war reporting unrelated to the German election campaign.
We therefore blacklisted these 10 keywords:
\textit{['btw25', 'btw2025', 'russland', 'russisch', 'ukrain', 'energie', 'wandel', 'wende', 'kohle', 'umwelt']}.
A post is excluded from the dataset only if \emph{all} of its matched keywords are blacklisted, so posts that mention these terms alongside any election-related keyword remain in the dataset.

\section{Opaque Research API Behaviors}
\label{sec:api-behaviors}
During the federal-election collection, we identified two undocumented behaviors of the TikTok research API.
We document them here as observed between January and April 2025.
The platform may change these behaviors at any time.

\paragraph{Trailing Digits Are Stripped From Keywords}
The research API silently ignores trailing digits in query keywords: a query for \textit{btw25} returns posts matching \textit{btw}, and \textit{wahl25} returns posts matching \textit{wahl}.
We discovered this when auditing which keyword each returned post had matched: thousands of posts returned for our query did not contain any of our keywords as defined.
Stripping trailing digits from the keywords explained 99.93\% of 64{,}538 such initially-unmatched posts.
This behavior has practical consequences: election-specific tags like \textit{btw25} (Bundestagswahl 2025) are common search terms in election research, and the API silently turns them into much broader queries, or even semantically unrelated ones, as with \textit{btw}.
The magnitude of the effect can be large: in the federal-election collection, 193{,}790 posts (26.1\%) contain the term \textit{wahl}, but only 28{,}683 of them (14.8\%) contain a digit-suffixed form such as \textit{wahl25}.
The keyword thus matched roughly seven times more posts than intended.
The Saxony collection is affected as well, through the keywords \textit{ltw24} and \textit{landtagswahl24}.

\paragraph{Random Sampling Draws From a Small Pool}
The research API offers a random-sampling mode (\textit{is\_random=true}) that is supposed to return a random sample of the posts matching a query.
Alongside our daily census collection, we also collected daily random samples.
Surprisingly, the random samples were much smaller than the censuses for the same query and day: repeated random-sample requests quickly returned only duplicates.
Capture--recapture estimates over the returned post IDs put the pool that the random mode samples from at roughly 200--560 posts per day, while our census collection for the same query and days retrieved 6{,}000--20{,}000 distinct posts (verified on 5 days spanning January--April 2025).
The random mode thus covers only \textasciitilde{}2--6\% of the matching population, and the inclusion criteria for this pool are opaque.
We conclude that the random-sampling mode was unsuitable for prevalence estimation during our collection period, and we do not release the random-sample subsets of our collections.

\section{Deletion Status Codes and Regimes}
\label{sec:removal-regimes}
When we re-scrape a post and it is no longer available, TikTok returns one or more status codes that describe why (for example \textit{status\_deleted}, \textit{status\_self\_see}, \textit{status\_reviewing}, \textit{status\_audit\_not\_pass}, \textit{cross\_border\_violation}, \textit{author\_status}).
A single post can carry several codes at once.
We group the codes into two regimes:

\begin{enumerate}
  \item \emph{Creator withdrawal:} a creator deleting a post, switching it to private or friends-only, or the account no longer existing.
  \item \emph{Platform enforcement:} a post under review, a failed moderation audit, age-restriction, geo-blocking, and the account-level status codes that accompany them.
\end{enumerate}

A deleted post is assigned to platform enforcement if it carries at least one enforcement code, and to creator withdrawal otherwise.

We separate the two regimes from the structure of the codes, not from their meaning alone.
The creator-withdrawal codes appear in isolation: \textit{status\_deleted} and \textit{status\_self\_see} are the only code on the post about 80\% of the time.
The enforcement codes almost never appear alone (under 1\%); they travel as a tightly correlated bundle.
For example, among posts under review, 87\% also carry the account-level status code \textit{author\_status}, and 99\% of geo-blocked posts also carry a failed-audit status code.
This structure is what lets us read the deletion mechanism off the codes without a deletion log.

Our platform-enforcement regime includes geo- and age-restriction, which TikTok's report (\autoref{sec:content-removal}) does not count as Community Guidelines removals. Only 1.3\% of posts carry such a code with no review or audit-failure code; excluding them lowers the rate from 13.0\% to 11.6\% (still 13--17$\times$).

How long after the election we re-scrape shapes both how much content is gone and why, as shown in \autoref{tab:removal-horizon}.
For Saxony we re-scraped the same 102{,}953 posts at three horizons (a panel: the same posts observed over time); both the deletion rate and the creator-withdrawal share among the deleted posts grow with the horizon.
The federal election was re-scraped once, 16 months after the election.
Its high deletion rate is consistent with the Saxony trend and reflects the longer observation horizon.
Platform enforcement is concentrated early, while creator withdrawals keep accumulating.

\begin{table}[h]
  \caption{\textbf{Deletions Over Time.} For each re-scrape: the share of re-scraped posts that were deleted, and, among those, the split between creator withdrawal and platform enforcement. The three Saxony rows are a panel of the same 102{,}953 posts re-scraped at 1, 3, and 4.5 months; the federal-election collection (743{,}588 posts) was re-scraped once at 16 months. Both the deletion rate and the creator-withdrawal share grow with the horizon.}
  \label{tab:removal-horizon}
  \centering
  \begin{tabular}{llrrr}
    \toprule
    Collection & Time after election & \% deleted & By Creator & By Platform \\
    \midrule
    Saxony 2024 (panel) & 1 month & 6.3\% & 51.5\% & 48.5\% \\
    Saxony 2024 (panel) & 3 months & 17.4\% & 54.8\% & 45.2\% \\
    Saxony 2024 (panel) & 4.5 months & 20.5\% & 55.9\% & 44.1\% \\
    Federal Election 2025 & 16 months & 39.7\% & 67.4\% & 32.6\% \\
    \bottomrule
  \end{tabular}
\end{table}

These deletion status codes are a snapshot at re-scrape, not a log of platform actions, so creator withdrawal cannot distinguish a creator deleting a single post from an account being deleted in full.
The status code \textit{author\_invalid} (an account that no longer exists) is counted as creator withdrawal but can also result from a ban, so the creator-withdrawal share is a mild upper bound.
Several status codes are undocumented by TikTok (for example \textit{author\_status} and \textit{author\_ftc}); we interpret them by their co-occurrence behavior rather than an official definition.

\section{Codebook}
\label{sec:all-codebooks}

\subsection{Politics}
\textit{Question:} Does the video refer to political topics (policy, politics or polity)?

\textit{Explanation:} content that discuss, reference, or engage with topics, events, figures, or movements that refer to policy, political events or societal challenges

\textit{Additional Information:} This might include:

\begin{itemize}
  \item news and current political events, e.g., elections
  \item political opinions and commentary by user, politicians, etc.
  \item activism and advocacy, e.g., promotion of social or political causes, such as environmental activism, LGBTQ+ rights, etc.
  \item political figures and campaigns, e.g., featuring or discussing political figures, candidates, parties, and campaigns
  \item satire and political memes
\end{itemize}

\subsection{Saxony}
\textit{Question:} Does the video refer to Saxony?

\textit{Explanation:} content that refers to the federal state of Saxony in any way

\textit{Additional Information:} This might include:

\begin{itemize}
  \item politics in Saxony, including the state election in Saxony
  \item politicians or celebrities from Saxony
  \item cities / landscape in Saxony
\end{itemize}

\subsection{Intolerance}
\label{sec:codebook-intolerance}

\textit{Question:} Does the video convey intolerance?

\textit{Explanation:} content that communicate the denial of an equal status with others and the refusal to accept different perspectives, cultures, or ideas

\textit{Additional Information:} This code is manifested through expressions of a harmful or discriminatory intent toward individuals or groups based on their social identities, preferences, or beliefs including

\begin{itemize}
  \item expressions of exclusion
  \item limiting people's (social groups) rights or undermining participation / silencing
  \item promotion of hostility towards what is seen as ``other'' or unfamiliar
  \item promotion of discriminatory or derogatory speech / symbols
\end{itemize}

\textit{Examples:}

\begin{itemize}
  \item Textual Elements: captions, on-screen text, or hashtags with language that is exclusionary and/or derogatory towards certain groups, people or viewpoints, e.g., dehumanizing insults, code words that express extreme ideology
  \item Visual Elements: symbols or depictions that mock or demean specific groups, including exaggerated, stereotypical visual cues, or discriminatory symbols that convey exclusion or marginalization, e.g., symbols / signs that express extreme ideology
  \item Audio Elements: tone, choice of words, or sound effects / music that convey hostility or mockery, including harsh, condescending tones, discriminatory language, or background audio that underscores a sense of judgment or rejection of certain individuals or groups
\end{itemize}

\subsection{Hedonic Entertainment}
\label{sec:codebook-hedonic-entertainment}
\textit{Question:} Does the video convey hedonic entertainment?

\textit{Explanation:} content that convey entertainment through enjoyment, amusement, and pleasure

\textit{Additional Information:} This code is manifested through hedonic entertainment cues that convey fun, humor, excitement, and pleasure (e.g., memes, sticker, emojis, music)

\textit{Examples:}

\begin{itemize}
  \item Textual Elements: Captions or on-screen text that use humor, slang, or lighthearted language aimed at amusement, such as jokes, punchlines, or exaggerated phrases; hashtags, emoticons or symbols that add a lighthearted or humorous tone
  \item Visual Elements: Bright colors, fast-paced cuts, and dynamic visual effects (e.g., sparkles, flashy transitions); Stickers, GIFs, Meme elements added to the video; playful facial expressions, animated gestures, or exaggerated movements that add humor or evoke joy.
  \item Audio Elements: Upbeat music, catchy soundtracks, or sound effects that evoke excitement, laughter, or surprise (e.g., cartoon sound); lively or energetic voices; laughter, audience reactions; telling jokes
\end{itemize}

\subsection{Eudaimonic Entertainment}
\label{sec:codebook-eudaimonic-entertainment}
\textit{Question:} Does the video convey eudaimonic entertainment?

\textit{Explanation:} content that convey entertainment through inspiring, touching, thought-provoking, and meaningful elements

\textit{Additional Information:} This code is manifested through eudaimonic entertainment cues that convey appreciation of beauty and excellence, gratitude, hope, personal growth, or religiousness/spirituality (e.g., visuals incl. nature or spiritual symbols, music)

\textit{Examples:}

\begin{itemize}
  \item Textual Elements: Thought-provoking captions, quotes, or questions that encourage introspection or convey moral messages/dilemma (e.g., "What truly matters in life?" or "Together, we are stronger"); hashtags or phrases that signal reflection, growth, gratitude or empathy
  \item Visual Elements: Soft or muted color tones, slower visual transitions, or minimalist visuals that focus on conveying depth, depiction of beauty and excellence; Imagery of personal stories or social issues, such as vulnerability or resilience; symbolic visuals, like sunsets or natural landscapes, architecture; references to life/death experiences, and religious and spiritual symbols
  \item Audio Elements: Calm or inspiring music, instrumental tracks, or soothing tones; slower-paced, gentle voice tones that emphasize reflection or encourage contemplation: sounds associated with nature or other tranquil settings (e.g., waves, birdsong)
\end{itemize}

\end{document}